\documentclass[11pt,a4paper]{article}
\usepackage{authblk}
\usepackage{amsmath,amsfonts,amssymb}
\usepackage{bbold}
\usepackage{bbm}
\usepackage{txfonts}
\usepackage[T1]{fontenc}
\usepackage{hyperref}
\def\bd{\begin{displaymath}}\def\ed{\end{displaymath}}
\def\be{\begin{equation}}\def\ee{\end{equation}}
\def\bea{\begin{eqnarray}}\def\eea{\end{eqnarray}}
\def\lb{\label}\def\bdot{\!\cdot\!}\def\ha{{1\over 2}}

\def\hx{\hat x}\def\hp{\hat p} 

\def\f{\varphi}\def\bp{\bar p}\def\bx{\bar x}

\newtheorem{proof}{Proof}

\numberwithin{equation}{section}
\numberwithin{theorem}{section}
\numberwithin{proposition}{section}
\numberwithin{remark}{section}
\numberwithin{example}{section}

\def\PL#1{\textit{Phys.\ Lett.}\ {\bf#1}}\def\CMP#1{Commun.\ Math.\ Phys.\ {\bf#1}}
\def\PRL#1{\textit{Phys.\ Rev.\ Lett.}\ {\bf#1}}
\def\PR#1{\textit{Phys.\ Rev.}\ {\bf#1}}\def\CQG#1{\textit{Class.\ Quantum Grav.}\ {\bf#1}}
\def\NP#1{\textit{Nucl.\ Phys.}\ {\bf#1}}
\def\JMP#1{\textit{J.\ Math.\ Phys.}\ {\bf#1}}

 \def\IJMP#1{\textit{Int.\ J. Mod.\ Phys.}\ {\bf #1}}
\def\MPL#1{\textit{Mod.\ Phys.\ Lett.}\ {\bf #1}}

\def\JHEP#1{\textit{JHEP}\ {\bf#1}}\def\JCAP#1{\textit{JCAP}\ {\bf#1}}
\def\RMP#1{\textit{Rev.\ Mod.\ Phys.}\ {\bf#1}}\def\AdP#1{Annalen Phys.\ {\bf#1}}
\def\arx#1{{\tt arXiv:#1}}

\newcommand{\bibp}[3]{#1, #2, #3}

\title{Generalized Triply Special Relativity models and their classical limit}

\begin{document}
\author[a]{Tea Martini\'{c} Bila\'{c}}
\affil[a]{Faculty of Science, University of Split, Rudjera Bo\v{s}kovi\'{c}a 33, 21000 Split, Croatia\\
	E-mail:\;\href{mailto:teamar@pmfst.hr}{teamar@pmfst.hr}}

\author[b]{Stjepan Meljanac}
\affil[b]{Rudjer Bo\v{s}kovi\'{c} Institute, Theoretical Physics Division, Bijeni\v{c}ka c. 54, HR 10002 Zagreb, Croatia\\
	E-mail:\;\href{mailto:meljanac@irb.hr}{meljanac@irb.hr}}

 \author[c]{Salvatore Mignemi}
 \affil[c]{Dipartimento di Matematica, Universit\`{a} di Cagliari via Ospedale 72, 09124 Cagliari, Italy and INFN, Sezione di Cagliari, Cittadella Universitaria, 09042 Monserrato, Italy\\
 	E-mail:\;\href{mailto:smignemi@unica.it}{smignemi@unica.it}}

\date{}
\maketitle

\begin{abstract}
Triply Special Relativity is a deformation of Special Relativity based on three fundamental parameters,
that describes a noncommutative geometry on a curved spacetime, preserving the Lorentz invariance and the
principle of relativity. Its symmetries are generated by a 14-parameter nonlinear algebra.

In this paper, we discuss a generalization of the original model and construct its realizations on a
canonical phase space.
We also investigate in more detail its classical limit, obtained by replacing the commutators by Poisson
brackets.

\end{abstract}

%%%%%%%%%%%%%%%%%%%%%%%%%%%%%%%%%%%%%%%%%%%%%%%%%%%%%%%%%%%%%%%%%%%%%
%%%%%%%%%%%%%%%%%%%%%%%%%%%%%%%%%%%%%%%%%%%%%%%%%%%%%%%%%%%%%%%%%%%%%

%%%%%%%%%%%%%%%%%%%%%%%%%%%%%%%%%%%%%%%%%%%%%%%%%%%%%%%%%%%%%%%%%%%
%%%%%%%%%%%%%%%%%%%%%%%%%%%%%%%%%%%%%%%%%%%%%%%%%%%%%%%%%%%%%%%%%%%
\section{Introduction}
Triply special relativity (TSR), also called Snyder-de Sitter (SdS) model, was proposed in 2004 as a
generalization of the Snyder model \cite{Snyder} of
noncommutative geometry \cite{ncg} to curved spacetime \cite{Kowalski}. Its name is due to the fact that it consists
in a deformation of doubly special relativity (DSR) \cite{dsr}, a theory based on two fundamental constants,
by a third parameter related to the curvature of spacetime.
The resulting theory therefore incorporates three deformations of Newtonian mechanics by three fundamental
coupling constants: special relativity, where the deformation is induced by the universality of
the speed of light $c$, DSR, with a deformation induced by the Planck mass $M_{Pl}$, and
a third one induced by a new length scale, related to the inverse of the cosmological constant $\Lambda$.
The theory is constructed in such a way to maintain the relativity principle
(all observers are equivalent) and the Lorentz invariance.

TSR presents some similarities with the Yang model \cite{Yang}, since it is based on the same symmetry
group $SO(1,5)$, generated by an algebra that includes position, momentum and Lorentz generators,
but is realized in a nonlinear way, so that one can get rid of the
fifteenth  generator of the $o(1,5)$ algebra, whose physical interpretation is not obvious.

The introduction of this model was suggested by considerations coming from quantum gravity,
where (at least in three dimensions), both the Planck mass and the cosmological constant
play a role in the low-energy limit \cite{Kowalski}.

Among the properties that characterize TSR is a duality that interchanges positions and momenta \cite{Guo}, generalizing
Born's old proposal \cite{Born}.
It must also be remarked that the physical properties of the model depend on the choice of the
sign of the deformation parameters. In particular, for positive signs of $\alpha^2$ and $\beta^2$, a maximal mass, given by $M_{Pl}$ and a
maximal length (radius of the universe) are predicted, while no such bound is present if the coupling constants
are negative. Notice that $\alpha^2$ and $\beta^2$ must have the same sign, to avoid complex structure constants.
For simplicity, in the following we consider the case of positive parameters.

In the past, the theory has been studied under several points of view, both in its original formulation and in
its classical limit, where commutators are replaced by Poisson brackets, obtaining a more manageable formalism.
Some relevant references for the classical framework are \cite{Mignemi-1}, where the realization of TSR by an
embedding in six-dimensional space was discussed, and \cite{Banerjee}, where a Hamiltonian formulation was given.
In a quantum setting, the nonrelativistic model was applied to some physical systems in \cite{Mignemi-2} and the
derivation of its representations from those of the Snyder model were discussed in \cite{Mignemi-2,Mignemi-3},
An attempt to the construction of a quantum field theory was also made in \cite{Franchino}.

In this paper, we consider nonlinear generalizations of the TSR model and their realizations on canonical phase space, first discussed in
\cite{Meljanac-1,Meljanac-2},
as a particular cases  of the models introduced in \cite{Meljanac-3}.

Realizations of the Yang algebra were also studied on an extended phase space, that included tensorial degrees of freedom \cite{Lukierski,Martinic-1},
but we shall not consider this possibility.

\section{Snyder model}
We start by shortly reviewing the Snyder model \cite{Snyder}, since many of the results valid in this case will be extended to TSR and its nonlinear
generalizations in the next sections.
We recall that it describes a noncommutative spacetime with a fundamental mass scale, but unbroken Lorentz invariance, exploiting a deformation
of the Heisenberg algebra by a parameter $\beta$.
Snyder's original formulation was generalized in \cite{Battisti} and the properties of its associated Hopf algebra were discussed in detail for example in
\cite{Meljanac-5}.

The Snyder model is generated by the noncommutative coordinates $\hx_\mu$ and the Lorentz generators $M_{\mu\nu}$, which obey the commutation relations
\bea\lb{cor}
&&[\hx_\mu,\hx_\nu]=i\beta^2 M_{\mu\nu},\qquad[M_{\mu\nu},\hx_\lambda]=i\left(\eta_{\mu\lambda}\hx_\nu-\eta_{\nu\lambda}\hx_\mu\right),\cr
&&[M_{\mu\nu},M_{\rho\sigma}]=i\big(\eta_{\mu\rho}M_{\nu\sigma}-\eta_{\mu\sigma}M_{\nu\rho}-\eta_{\nu\rho}M_{\mu\sigma}+\eta_{\nu\sigma}M_{\mu\rho}\big),
\eea
where $\beta$ is a constant of the order of $1/M_{Pl}$, and $\eta_{\mu\nu}=$\ diag $(-1,1,1,1)$.

%\qquad,f_1+\b^2p_\m p_\n\f_2),\eqno(1)
The Heisenberg algebra is generated by coordinates $x_\mu$ and momenta $p_\mu$, that satisfy
\be
[x_\mu,x_\nu]=[p_\mu,p_\nu]=0,\qquad[x_\mu,p_\nu]=i\eta_{\mu\nu},
\ee
transforming as vectors under the action of the Lorentz algebra
\be
[M_{\mu\nu},x_\lambda]=i\left(\eta_{\mu\lambda}x_\nu-\eta_{\nu\lambda}x_\mu\right),\qquad[M_{\mu\nu},p_\lambda]=i\left(\eta_{\mu\lambda}p_\nu-\eta_{\nu\lambda}p_\mu\right).
\ee

General realizations of the Snyder model in terms of the Heisenberg algebra, linear in the coordinates $x_\mu$, are \cite{Battisti,Meljanac-5,Martinic-2}
\be
\hx_\mu=x_\mu \f_1(u)+\beta^2 \,x\bdot p\,p_\mu\f_2(u),
\ee
with $u=\beta^2p^2$ and $\f_1(0)=1$. The Jacobi identities are satisfied if
\be \label{s-j}
\f_2={1+2\f_1\f_1'\over\f_1-2u\f_1'},\qquad{\rm with}\quad\f_1'={d\f_1\over du}.
\ee
and
\be
M_{\mu\nu}=x_\mu p_\nu-x_\nu p_\mu.
\ee
It follows that
\be
[\hx_\mu,p_\nu]=i(\eta_{\mu\nu}\f_1+\beta^2p_\mu p_\nu\f_2).
\ee
The original Snyder realization is obtained for $\f_1(u)=\f_2(u)=1$ and the important realization found by
Maggiore \cite{Maggiore} for $\f_1(u)=\sqrt{1-u}$, $\f_2(u)=0$.

\section{Generalized TSR models}

\bigskip
We pass now to define the generalized TSR models and their representations. These models were introduced in \cite{Meljanac-1,Meljanac-2}.

Let us define new noncommutative momenta as
\be\lb{def}
\hp_\mu=p_\mu-{\alpha\over\beta}\hx_\mu,
\ee
where $\alpha$ is a parameter of dimension $1/L$, whose square may be identified with the cosmological constant.
They satisfy
\be\lb{corp}
[\hp_\mu,\hp_\nu]=i\alpha^2M_{\mu\nu},\qquad[M_{\mu\nu},\hp_\lambda]=i\left(\eta_{\mu\lambda}\hp_\nu-\eta_{\nu\lambda}\hp_\mu\right),
\ee
\be \label{s-1}
[\hx_\mu,\hp_\nu]=i\left(\eta_{\mu\nu}\f_1+(\alpha^2\hx_\mu\hx_\nu+\beta^2\hp_\mu\hp_\nu+\alpha\beta\hx_\mu\hp_\nu+\alpha\beta\hp_\mu\hx_\nu)\f_2-\alpha\beta M_{\mu\nu}\right),
\ee
where now $\f_i=\f_i(\alpha^2\hx^2+\beta^2\hp^2+\alpha\beta\hx\bdot\hp+\alpha\beta\hp\bdot\hx)$, $i=1,2$.
From (\ref{def}) also follows that $\beta p_\mu=\alpha\hx_\mu+\beta\hp_\mu$.

Note that algebra generated by $\hat{x}_{\mu}, \; \hat{p}_{\mu}$ and $M_{\mu\nu}$ defined in \eqref{cor}, \eqref{corp} and \eqref{s-1} is invariant under similarity transformations $\hat{x}_{\mu}\rightarrow S \hat{x}_{\mu} S^{-1}, \; \hat{p}_{\mu}\rightarrow S \hat{p}_{\mu} S^{-1}$ and $M_{\mu\nu}\rightarrow S M_{\mu\nu} S^{-1}$. If $\left[ M_{\mu\nu}, S\right] =0$ then $M_{\mu\nu}\rightarrow  M_{\mu\nu}$.

Furthermore,
\bea\lb{Mmn}
M_{\mu\nu}&=&x_\mu p_\nu-x_\nu p_\mu=\left(\hx_\mu\hp_\nu-\hx_\nu\hp_\mu\right){1\over\f_1-i\alpha\beta}\cr
&=&\ha\left(\hx_\mu\hp_\nu+\hp_\nu\hx_\mu-\hx_\nu\hp_\mu-\hp_\mu\hx_\nu\right){1\over\f_1}.
\eea
Note that $[M_{\mu\nu},\f_i(u)]=0$, and using (\ref{Mmn}) we can write
\bea\lb{corpx}
[\hx_\mu,\hp_\nu]&=&i\bigg(\eta_{\mu\nu}\f_1+(\alpha^2\hx_\mu\hx_\nu+\beta^2\hp_\mu\hp_\nu+\alpha\beta\hx_\mu\hp_\nu+\alpha\beta\hp_\mu\hx_\nu)\f_2\cr
&&-{\alpha\beta\over2}\left(\hx_\mu\hp_\nu+\hp_\nu\hx_\mu-\hx_\nu\hp_\mu-\hp_\mu\hx_\nu\right){1\over\f_1}\bigg).
\eea
For $\f_1=\f_2=1$, we get
\bea
[\hx_\mu,\hp_\nu]&=&i\Big(\eta_{\mu\nu}+\alpha^2\hx_\mu\hx_\nu+\beta^2\hp_\mu\hp_\nu+\alpha\beta\hx_\mu\hp_\nu+\alpha\beta\hp_\mu\hx_\nu\Big)-\alpha\beta M_{\mu\nu}\Big)\cr
&&=i\Big(\eta_{\mu\nu}+\alpha^2\hx_\mu\hx_\nu+\beta^2\hp_\nu\hp_\mu+\alpha\beta\hx_\nu\hp_\mu+\alpha\beta\hp_\mu\hx_\nu)\Big),
\eea
that corresponds to the original TSR model.

For $\f_1=\sqrt{1-\beta^2p^2}$, $\f_2=0$, we get
\be
[\hx_\mu,\hp_\nu]=i\eta_{\mu\nu}\sqrt{1-(\alpha^2\hx^2+\beta^2\hp^2+\alpha\beta\hx\bdot\hp+\alpha\beta\hp\bdot\hx)}-\alpha\beta M_{\mu\nu}
\ee
that corresponds to a different model of TSR, i.e.~a nonlinear algebra different from the algebra quadratic
in $\hx_\mu$ and $\hp_\mu$.
Defining $h=\sqrt{1-(\alpha\hx_\mu+\beta\hp_\mu)^2}$, one can write in this case,
\bea
&&\qquad\qquad\qquad[\hx_\mu,\hp_\nu]=i\eta_{\mu\nu} h-\alpha\beta M_{\mu\nu},\cr
&&[h,\hx_\mu]=\beta(\alpha\hx_\mu+\beta\hp_\mu),\qquad[h,\hp_\mu]=-\alpha(\alpha\hx_\mu+\beta\hp_\mu).
\eea

Hence, we have constructed a family of new nonlinear algebras corresponding to generalized TSR models different
from the original one.

An alternative dual construction is to start from the de Sitter algebra
\be
[\hp_\mu,\hp_\nu]=i\alpha^2M_{\mu\nu}
\ee
and define noncommutative coordinates
\be
\hx_\mu=x_\mu-{\beta\over\alpha}\hp_\mu
\ee
with dual realizations of $\hp_\mu$,
\be
\hp_\mu=p_\mu \f_1(v)+\alpha^2 \,p\bdot x\,x_\mu\f_2(v),
\ee
with $v=\alpha^2x^2$.

One can also define Hermitian realizations for $\hx_\mu$ and $\hp_\mu$ obtained simply by
\be
\hx_\mu\to\ha\left(\hx_\mu+\hx_\mu^\dagger\right),\qquad\hp_\mu\to\ha\left(\hp_\mu+\hp_\mu^\dagger\right).
\ee

To summarize, a realization of the generalized TSR model satisfying (\ref{cor}), (\ref{corp}) and (\ref{corpx}) is given by
\bea\lb{realp}
\hx_\mu&=&x_\mu \f_1(u)+\beta^2 \,x\bdot p\,p_\mu\f_2(u),\cr
\hp_\mu&=&p_\mu-{\alpha\over\beta}\left(x_\mu \f_1(u)+\beta^2 \,x\bdot p\,p_\mu\f_2(u)\right),
\eea
and its dual realization is
\bea\lb{realx}
\hp_\mu&=&p_\mu \f_1(v)+\alpha^2p\bdot xx_\mu\,\f_2(v),\cr
\hx_\mu&=&x_\mu-{\beta\over\alpha}\left(p_\mu \f_1(v)+\alpha^2 \,p\bdot x\,x_\mu\f_2(v)\right).
\eea

A family of realizations interpolating between (\ref{realp}) and (\ref{realx}) is given by
\bea
\hx_\mu&=&\bx_\mu \f_1(\beta^2\bp^2)+\beta^2 \bx\bdot \bp\bp_\mu\f_2(\beta^2\bp^2)+{\beta\over\alpha}a\,\bp_\mu,\cr
\hp_\mu&=&(1-a)\bp_\mu-{\alpha\over\beta}\left(\bx_\mu \f_1(\beta^2\bp^2)+\beta^2 \,\bx\bdot\bp\,\bp_\mu\f_2(\beta^2\bp^2)\right),
\eea
where
\be
\bx_\mu=x_\mu\cos\epsilon-{\beta\over\alpha}p_\mu\sin\epsilon,\qquad\bp_\mu=p_\mu\cos\epsilon+{\alpha\over\beta}x_\mu\sin\epsilon.
\ee
For $\epsilon=0$, $a=0$, we get eq.~(\ref{realp}), and for $\epsilon={\pi\over2}$, $a=1$ the dual realization, eq.~(\ref{realx}).
For $\epsilon={\pi\over4}$, $a=\ha$ we get symmetric realizations,
\be
\hp_\mu=\hx_\mu\Big|_{x\leftrightarrow p},\qquad\hx_\mu=\hp_\mu\Big|_{x\leftrightarrow p}.
\ee

Corresponding Hermitian realizations are obtained in all cases changing $\hx_\mu\to\ha\left(\hx_\mu+\hx_\mu^\dagger\right)$
and $\hp_\mu\to\ha\left(\hp_\mu+\hp_\mu^\dagger\right)$.
For the original TSR model another symmetric realization is obtained in sect.~5 of \cite{Meljanac-1}, see also \cite{Meljanac-2}.

The algebra generated by $\hx_\mu$ and $\hp_\mu$ and $M_{\mu\nu}$ and all its realizations are invariant under Born duality,
\be
\alpha\to-\beta,\quad\beta\to\alpha,\quad\hx_\mu\to-\hp_\mu,\quad\hp_\mu\to\hx_\mu,\quad M_{\mu\nu}\leftrightarrow M_{\mu\nu}.
%\quad g_\mn\lra g_\mn
\ee

We also recall that a different algebra can be obtained if both $\alpha^2$ and $\beta^2$ are negative. 
However, contrary for example to the Yang model, $\alpha^2$ and $\beta^2$ cannot have opposite signs.

For a particular realization of a given algebra, all other realizations can be obtained by similarity transformations,
see \cite{Martinic-2}.

\section{Generalized TSR Poisson models}

The generalized TSR models introduced above can be investigated in their classical limit, where the commutators are replaced by Poisson brackets.
This is similar to what was done in \cite{Meljanac-4,Martinic-3} for the Yang model and allows for a simpler investigation, avoiding problems related to the ordering of the operators.

The TSR Poisson model is generated with $\hat{x}_{\mu}, \; \hat{p}_{\mu}$ and Lorentz generators $M_{\mu\nu}$,
\begin{equation}\label{0.1}
\left\lbrace M_{\mu\nu},M_{\rho\sigma}\right\rbrace =\left( \eta_{\mu\rho}M_{\nu\sigma}-\eta_{\mu\sigma}M_{\nu\rho}-\eta_{\nu\rho}M_{\mu\sigma}+\eta_{\nu\sigma}M_{\mu\rho}\right),
\end{equation}
\begin{equation}\label{0.2}
\left\lbrace M_{\mu\nu}, \hat{x}_{\lambda}\right\rbrace  =\left( \eta_{\mu\lambda}\hat{x}_{\nu}-\eta_{\nu\lambda}\hat{x}_{\mu}\right) ,\quad  \left\lbrace M_{\mu\nu}, \hat{p}_{\lambda}\right\rbrace  =\left( \eta_{\mu\lambda}\hat{p}_{\nu}-\eta_{\nu\lambda}\hat{p}_{\mu}\right)
\end{equation}
\begin{equation}\label{0.3}
\left\lbrace \hat{x}_{\mu},\hat{x}_{\nu}\right\rbrace =\beta^{2}M_{\mu\nu}, \quad \left\lbrace \hat{p}_{\mu},\hat{p}_{\nu}\right\rbrace =\alpha^{2}M_{\mu\nu},
\end{equation}
\begin{equation}\label{0.4}
\left\lbrace \hat{x}_{\mu}, \hat{p}_{\nu}\right\rbrace =\eta_{\mu\nu}\varphi_{1}(u)+\left( \alpha \hat{x}_{\mu}+\beta\hat{p}_{\mu}\right)\left( \alpha \hat{x}_{\nu}+\beta\hat{p}_{\nu}\right)\varphi_{2}(u) -\alpha\beta M_{\mu\nu},
\end{equation}
where  $\; \eta_{\mu\nu}=diag(-1,1,1,1)$ is the Minkowski metric, $\alpha$ and $\beta$ are real parameters and $u=\beta^{2}p^{2}$.
The TSR Poisson model is constructed from \eqref{cor}, \eqref{corp} and \eqref{s-1} where we replace commutators $\left[\ ,\ \right] $ with classical Poisson brackets $\left\lbrace\ ,\ \right\rbrace $ and we remove the factor $i$. The Jacobi identities for the Poisson brackets are satisfied if and only if \eqref{s-j} holds.

\bigskip

We look for realizations of $\hat{x}_{\mu}$ and $\hat{p}_{\mu}$ on a phase space with coordinates $x_{\mu}$ and momenta $p_{\mu}$ satisfying the canonical algebra
\begin{equation}
\left\lbrace x_{\mu},x_{\nu}\right\rbrace =\left\lbrace p_{\mu},p_{\nu}\right\rbrace =0, \quad \left\lbrace x_{\mu},p_{\nu}\right\rbrace =\eta_{\mu\nu}.
\end{equation}

The general ansatz for $\hat{x}_{\mu}\;$ and $\hat{p}_{\mu}$ is
\begin{equation}\label{0.6}
\hat{x}_{\mu}=x_{\mu}f+\frac{\beta}{\alpha}p_{\mu}g
\end{equation}
and
\begin{equation}\label{0.7}
\hat{p}_{\mu}=p_{\mu}\tilde{f}+\frac{\alpha}{\beta}x_{\mu}\tilde{g},
\end{equation}
where $f,g,\tilde{f},\tilde{g}\;$ are functions of $u,v,z\;$ and $\;u=\beta^{2}p^{2}, \; v=\alpha^{2}x^{2}, \; z=\alpha\beta x\bdot p$.\\
We find corresponding differential equations for $f,g,\tilde{f},\tilde{g}\;$.
From $\left\lbrace \hat{x}_{\mu}, \hat{x}_{\nu}\right\rbrace = \beta^{2}M_{\mu\nu}$ we get
\begin{align}
&-2f\frac{\partial f}{\partial u}-2g\frac{\partial g}{\partial v}+4z\left( \frac{\partial f}{\partial v}\frac{\partial g}{\partial u}-
\frac{\partial f}{\partial u}\frac{\partial g}{\partial v}\right) +2v\left( \frac{\partial f}{\partial v}\frac{\partial g}{\partial z}-\frac{\partial f}{\partial z}\frac{\partial g}{\partial v}\right) \notag\\
&+2u\left( \frac{\partial f}{\partial z}\frac{\partial g}{\partial u}-\frac{\partial f}{\partial u}\frac{\partial g}{\partial z}\right)+f\frac{\partial g}{\partial z}+g\frac{\partial f}{\partial z}
=1 \label{e-1}
\end{align}
and from $\left\lbrace \hat{p}_{\mu}, \hat{p}_{\nu}\right\rbrace =\alpha^{2}M_{\mu\nu}$ it follows
\begin{align}
&-2\tilde{f}\frac{\partial \tilde{f}}{\partial v}-2\tilde{g}\frac{\partial \tilde{g}}{\partial u}+4z\left( \frac{\partial \tilde{f}}{\partial u}\frac{\partial \tilde{g}}{\partial v}-
\frac{\partial \tilde{f}}{\partial v}\frac{\partial \tilde{g}}{\partial u}\right) +2v\left( \frac{\partial \tilde{f}}{\partial z}\frac{\partial \tilde{g}}{\partial v}-\frac{\partial \tilde{f}}{\partial v}\frac{\partial \tilde{g}}{\partial z}\right) \notag\\
&+2u\left( \frac{\partial \tilde{f}}{\partial u}\frac{\partial \tilde{g}}{\partial z}-\frac{\partial \tilde{f}}{\partial z}\frac{\partial \tilde{g}}{\partial u}\right)+\tilde{f}\frac{\partial \tilde{g}}{\partial z}+\tilde{g}\frac{\partial \tilde{f}}{\partial z}
=1.\label{e-2}
\end{align}
The relation \eqref{0.4} yields following five equations:
\begin{equation} \label{e-3}
f\tilde{f}-g\tilde{g}=\varphi_{1}(u),
\end{equation}
\begin{align}
&2\tilde{f}\frac{\partial f}{\partial v}-2g\frac{\partial \tilde{g}}{\partial v}+4z\left( \frac{\partial f}{\partial v}\frac{\partial \tilde{g}}{\partial u}-
\frac{\partial f}{\partial u}\frac{\partial \tilde{g}}{\partial v}\right) +2v\left( \frac{\partial f}{\partial v}\frac{\partial \tilde{g}}{\partial z}-\frac{\partial f}{\partial z}\frac{\partial \tilde{g}}{\partial v}\right) \notag \\
&+2u\left( \frac{\partial f}{\partial z}\frac{\partial \tilde{g}}{\partial u}-\frac{\partial f}{\partial u}\frac{\partial \tilde{g}}{\partial z}\right)+f\frac{\partial \tilde{g}}{\partial z}-\tilde{g}\frac{\partial f}{\partial z}
=\left( f+\tilde{g}\right) ^{2}\varphi_{2}(u),\label{e-4}
\end{align}
\begin{align}
&-2f\frac{\partial \tilde{f}}{\partial u}-2\tilde{g}\frac{\partial g}{\partial u}+4z\left( \frac{\partial \tilde{f}}{\partial u}\frac{\partial g}{\partial v}-
\frac{\partial \tilde{f}}{\partial v}\frac{\partial g}{\partial u}\right) +2v\left( \frac{\partial \tilde{f}}{\partial z}\frac{\partial g}{\partial v}-\frac{\partial \tilde{f}}{\partial v}\frac{\partial g}{\partial z}\right) \notag \\
&+2u\left( \frac{\partial \tilde{f}}{\partial u}\frac{\partial g}{\partial z}-\frac{\partial \tilde{f}}{\partial z}\frac{\partial g}{\partial u}\right)+\tilde{f}\frac{\partial g}{\partial z}-g\frac{\partial \tilde{f}}{\partial z}=\left( \tilde{f}+g\right) ^{2}\varphi_{2}(u),\label{e-5}
\end{align}
\begin{align}
&-2\tilde{g}\frac{\partial f}{\partial u}-2g\frac{\partial \tilde{f}}{\partial v}+4z\left( \frac{\partial f}{\partial v}\frac{\partial \tilde{f}}{\partial u}-
\frac{\partial f}{\partial u}\frac{\partial \tilde{f}}{\partial v}\right) +2v\left( \frac{\partial f}{\partial v}\frac{\partial \tilde{f}}{\partial z}-\frac{\partial f}{\partial z}\frac{\partial \tilde{f}}{\partial v}\right)\notag\\
&+2u\left( \frac{\partial f}{\partial z}\frac{\partial \tilde{f}}{\partial u}-\frac{\partial f}{\partial u}\frac{\partial \tilde{f}}{\partial z}\right)+f\frac{\partial \tilde{f}}{\partial z}+\tilde{f}\frac{\partial f}{\partial z}
=\left( f+\tilde{g}\right) \left( \tilde{f}+g\right)\varphi_{2}(u)  -1,\label{e-6}
\end{align}
and
\begin{align}
&2f\frac{\partial \tilde{g}}{\partial u}+2\tilde{f}\frac{\partial g}{\partial v}+4z\left( \frac{\partial g}{\partial v}\frac{\partial \tilde{g}}{\partial u}-
\frac{\partial g}{\partial u}\frac{\partial \tilde{g}}{\partial v}\right) +2v\left( \frac{\partial g}{\partial v}\frac{\partial \tilde{g}}{\partial z}-\frac{\partial g}{\partial z}\frac{\partial \tilde{g}}{\partial v}\right) \notag \\
&+2u\left( \frac{\partial g}{\partial z}\frac{\partial \tilde{g}}{\partial u}-\frac{\partial g}{\partial u}\frac{\partial \tilde{g}}{\partial z}\right)-g\frac{\partial \tilde{g}}{\partial z}-\tilde{g}\frac{\partial g}{\partial z}
=\left( f+\tilde{g}\right) \left( \tilde{f}+g\right) \varphi_{2}(u)+1.\label{e-7}
\end{align}

In the limit when $\alpha \rightarrow 0$, the algebra \eqref{0.1}-\eqref{0.4} becomes the Snyder Poisson algebra. Choosing $\tilde{f}=1,\; \tilde{g}=0$, i.e.~$\hat{p}_{\mu}=p_{\mu}$ we get a realization of $\hat{x}_{\mu}$
\begin{equation}
\hat{x}_{\mu}=x_{\mu}\varphi_{1}(u)+\frac{\beta}{\alpha}p_{\mu}z\varphi_{2}(u),
\end{equation}
where $\varphi_{2}(u)$ satisfies \eqref{s-j}.
Analogously in the limit when $\beta \rightarrow 0$, the algebra \eqref{0.1}-\eqref{0.4} becomes the dual Snyder Poisson algebra.
Choosing $f=1,\; g=0$, i.e.~$\hat{x}_{\mu}=x_{\mu}$ we get a realization of $\hat{p}_{\mu}$
\begin{equation}
\hat{p}_{\mu}=p_{\mu}\varphi_{1}(v)+\frac{\alpha}{\beta}x_{\mu}z\varphi_{2}(v),
\end{equation}
where $\varphi_{2}(v)$ satisfies \eqref{s-j} with $u\leftrightarrow v$.

It is important to note that \eqref{e-1} is invariant under the interchange
$f \leftrightarrow g$, $u \leftrightarrow v.$
This means that if $\hat {x}_{\mu} = x_{\mu} f + \frac{\beta}{\alpha} p_{\mu} g$ is a realization satisfying \eqref{e-1}, then
$\hat x_{\mu} =  x_{\mu} g\mid_{u \leftrightarrow v}  + \frac{\beta}{\alpha} p_{\mu} f\mid_{u \leftrightarrow v}$ is also a realization satisfying \eqref{e-1}.
The same holds in \eqref{e-2} with $\tilde {f} \leftrightarrow \tilde {g}, u \leftrightarrow v$.
This applies to \eqref{e-3}-\eqref{e-7}, but up to opposite $\pm$ sign.
It holds that \eqref{e-1} and \eqref{e-2} are related to each other by $f \leftrightarrow \tilde{f},\; g \leftrightarrow \tilde{g},\; u \leftrightarrow v$.

From the realization
\begin{equation}\label{1}
\hat{x}_{\mu}=x_{\mu}\varphi_{1}(u)+\frac{\beta}{\alpha}p_{\mu}z\varphi_{2}(u)
\end{equation}
it follows
\begin{equation}
f=\varphi_{1}(u), \quad g=z\varphi_{2}(u)
\end{equation}
and
\begin{equation}
\tilde{f}=1-z\varphi_{2}(u), \quad \tilde{g}=-\varphi_{1}(u).
\end{equation}

From the dual realization
\begin{equation}\label{1.1}
\hat{p}_{\mu}=p_{\mu}\varphi_{1}(v)+\frac{\alpha}{\beta}x_{\mu}z\varphi_{2}(v)
\end{equation}
it follows
\begin{equation}
\tilde{f}=\varphi_{1}(v), \quad \tilde{g}=z\varphi_{2}(v)
\end{equation}
and
\begin{equation}
f=1-z\varphi_{2}(v), \quad g=-\varphi_{1}(v).
\end{equation}
\bigskip

In particular, for $\varphi_{1}(u)=\varphi_{2}(u)=1$ it follows
\begin{equation}
f=1, \quad g=z
\end{equation}
and
\begin{equation}
\tilde{f}=1-z, \quad \tilde{g}=-1.
\end{equation}

For the dual realization $\varphi_{1}(v)=\varphi_{2}(v)=1$ it follows

\begin{equation}
\tilde{f}=1, \quad \tilde{g}=z
\end{equation}
and
\begin{equation}
f=1-z, \quad g=-1.
\end{equation}
\bigskip

A second special case is obtained from $\varphi_{1}(u)=\sqrt{1-u},\; \varphi_{2}(u)=0$, where
\begin{equation}
f=\sqrt{1-u}, \quad g=0
\end{equation}
and
\begin{equation}
\tilde{f}=1, \quad \tilde{g}=-\sqrt{1-u}.
\end{equation}
The dual realization $\varphi_{1}(v)=\sqrt{1-v},\; \varphi_{2}(v)=0$, gives
\begin{equation}
\tilde{f}=\sqrt{1-v}, \quad \tilde{g}=0,
\end{equation}
and
\begin{equation}
f=1, \quad g=-\sqrt{1-v}.
\end{equation}
\bigskip

Besides the above solutions there are self dual solutions $\tilde{f}(u,v,z)=f(v,u,z)\;$ and $\tilde{g}(u,v,z)=g(v,u,z)\;$ intherchanging $u\leftrightarrow v$. We consider the special case where
$\varphi_{1}(u)=\varphi_{2}(u)=1$. Let $f=1+\sum_{n=1}^{\infty}P_{n},\; g=\sum_{n=1}^{\infty}R_{n},\; \tilde{f}=1+\sum_{n=1}^{\infty}\tilde{P}_{n}\;$ and $\tilde{g}=\sum_{n=1}^{\infty}\tilde{R}_{n},$
with
\begin{equation}
P_{n}=\sum_{k+l+m=n}c_{klm}u^{k}v^{l}z^{m}, \quad R_{n}=\sum_{k+l+m=n}d_{klm}u^{k}v^{l}z^{m}
\end{equation}
and
\begin{equation}
\tilde{P}_{n}=\sum_{k+l+m=n}\tilde{c}_{klm}u^{k}v^{l}z^{m}, \quad \tilde{R}_{n}=\sum_{k+l+m=n}\tilde{d}_{klm}u^{k}v^{l}z^{m}.
\end{equation}
Using equations $\eqref{e-1}-\eqref{e-7}$ we find solutions for $f,g,\tilde{f}$ and $\tilde{g}$ up to second order:
\begin{align*}
f&=1+au-av+tu^{2}+(a^{2}+\frac{1}{2}b-t)v^{2}+\frac{(1+2a)^{2}}{2}z^{2}+\left( \frac{1}{8}+\frac{1}{2}b^{2}-a^{2}\right) uv\\
&+\left( s-2a-2ab-\frac{1}{2}\right) uz+\left( 4ab+3a+b+1-s\right) vz,
\end{align*}
\begin{align*}
g&=bu+\frac{1}{2}v+\left( 1+2a\right) z+ wu^{2}+rv^{2}+sz^{2}+\left( -\frac{1}{2}a+2b+6ab-2r\right) uv\\
&+\left( a+b+4a^{2}+4t\right) uz+\left( \frac{3}{4}+b^{2}+2a^{2}+3a\right) vz,
\end{align*}
\begin{align*}
\tilde{f}&=1-au+av+tv^{2}+(a^{2}+\frac{1}{2}b-t)u^{2}+\frac{(1+2a)^{2}}{2}z^{2}+\left( \frac{1}{8}+\frac{1}{2}b^{2}-a^{2}\right) uv\\
&+\left( s-2a-2ab-\frac{1}{2}\right) vz+\left( 4ab+3a+b+1-s\right) uz,
\end{align*}
\begin{align*}
\tilde{g}&=bv+\frac{1}{2}u+\left( 1+2a\right) z+ ru^{2}+wv^{2}+sz^{2}+\left( -\frac{1}{2}a+2b+6ab-2r\right) uv\\
&+\left( a+b+4a^{2}+4t\right) vz+\left( \frac{3}{4}+b^{2}+2a^{2}+3a\right) uz,
\end{align*}
where $a,b,r,s,t,w$ are real parameters. Note that for the choice of parameters $a=b=t=r=w=0\;$ and $\;s=1$ we get the solution
\begin{equation*}
f=1+\frac{uv}{8}+\frac{z^{2}}{2}+\frac{uz}{2}, \quad g=z+\frac{v}{2}+z^{2}+\frac{3}{4}vz,
\end{equation*}
\begin{equation*}
\tilde{f}=1+\frac{uv}{8}+\frac{z^{2}}{2}+\frac{vz}{2}, \quad \tilde{g}=z+\frac{u}{2}+z^{2}+\frac{3}{4}uz.
\end{equation*}
which is related to the realization
\begin{align}\notag
&\hat{x}_{\mu}=x_{\mu}+  \frac{\beta^{2}}{2}p_{\mu}p\bdot x + \frac{\alpha\beta}{4}p_{\mu}x^{2}+\frac{\alpha^{2}\beta^{2}}{16}x_{\mu}x^{2}p^{2}+\frac{\alpha^{2}\beta^{2}}{4}x_{\mu}x\bdot p p\bdot x +\frac{\alpha\beta^{3}}{4}x_{\mu}x\bdot p p^{2}\\
&+\frac{\alpha\beta^{3}}{2}p_{\mu}p \bdot x x \bdot p + \frac{3\alpha^{2}\beta^{2}}{8}p_{\mu}x^{2}x\bdot p + \text{h.c.},
\end{align}
\begin{align}\notag
&\hat{p}_{\mu}=p_{\mu}+  \frac{\alpha^{2}}{2}x_{\mu}x\bdot p + \frac{\alpha\beta}{4}x_{\mu}p^{2}+\frac{\alpha^{2}\beta^{2}}{16}p_{\mu}p^{2}x^{2}+\frac{\alpha^{2}\beta^{2}}{4}p_{\mu}p\bdot x x\bdot p +\frac{\alpha^{3}\beta}{4}p_{\mu}p\bdot x x^{2}\\
&+\frac{\alpha^{3}\beta}{2}x_{\mu}x \bdot p p \bdot x  +\frac{3\alpha^{2}\beta^{2}}{8}x_{\mu}p^{2}p\bdot x + \text{h.c.},
\end{align}
in \cite{Meljanac-1}, but with commuting $x$ and $p$, i.e.~$\left[ x,p\right] =0$.
\bigskip

We also find a family of realizations interpolating between \eqref{1} and $\eqref{1.1}$, which are given by
\begin{equation}
\hat{x}_{\mu}= \bar{x}_{\mu}\varphi_{1}(\beta^{2}\bar{p}^{2})+\beta^{2}\bar{p}_{\mu}\left( \bar{x}\bar{p}\right) \varphi_{2}(\beta^{2}\bar{p}^{2})
+\frac{\beta}{\alpha}a\bar{p}_{\mu}
\end{equation}
and
\begin{equation}
\hat{p}_{\mu}=\left( 1-a\right) \bar{p}_{\mu} -\frac{\alpha}{\beta}\left( \bar{x}_{\mu}\varphi_{1}(\beta^{2}\bar{p}^{2})+\beta^{2}\bar{p}_{\mu}\left( \bar{x}\bar{p}\right) \varphi_{2}(\beta^{2}\bar{p}^{2})
\right),
\end{equation}
where
\begin{equation}\label{1.2}
\bar{x}_{\mu}=x_{\mu}\cos \epsilon - \frac{\beta}{\alpha}p_{\mu} \sin \epsilon, \quad  \bar{p}_{\mu}=p_{\mu}\cos \epsilon + \frac{\alpha}{\beta}x_{\mu} \sin \epsilon.
\end{equation}

If we fix $\varphi_{1}(u)$ and $\varphi_{2}(u)$ in \eqref{e-3}-\eqref{e-7}, then for any solution for $f,g, \tilde{f}, \tilde{g}$ of \eqref{e-1}-\eqref{e-7}, a new family of solutions is generated by transformation in $\hat{x}_{\mu},\; \hat{p}_{\mu}$ \eqref{0.6}, \eqref{0.7} given by $x_{\mu}\rightarrow \bar{x}_{\mu}, \; p_{\mu}\rightarrow \bar{p}_{\mu}$, where  $\bar{x}_{\mu}$ and $\bar{p}_{\mu}$ are defined in \eqref{1.2}.

There are also more general solutions, for example for $f=1$ from \eqref{e-1} we get
\begin{equation}
-2g\frac{\partial g}{\partial v}+\frac{\partial g}{\partial z}=1
\end{equation}
which yields new additional solution $g(u,v,z)= \frac{1}{2}\left( z-\sqrt{1-2v-z^{2}}\right) $.

Let us now expand $g$ up to second order in $v,z$. Then we have $g=-\frac{1}{2}+\frac{1}{2}\left( v+z\right) +\frac{1}{4}\left( v^{2}+z^{2}\right) $. If we denote $\tilde{g}=a_{0}+a_{1}u+a_{2}v+a_{3}z+a_{4}u^{2}+a_{5}v^{2}+a_{6}z^{2}
+a_{7}uv+a_{8}uz+a_{9}vz$, then from \eqref{e-3} we have $\tilde{f}=1+g\tilde{g}$. Now, from $\eqref{e-1}-\eqref{e-7}$ we find
that
\begin{align}
& a_{2}+a_{3}=\left( 1+a_{0}\right) ^{2}, \quad a_{7}+a_{8}=2a_{1}+2a_{0}a_{1}, \label{c-1} \\
&a_{9}+2a_{5}-a_{2}=2a_{2}+2a_{0}a_{2}, \quad a_{9}+2a_{6}-a_{2}=2a_{3}+2a_{0}a_{3},\\
&2a_\textbf{1}+a_{3}+\frac{1}{2}a_{2}=1+2a_{0}, \quad 4a_{4}+\frac{1}{2}a_{8}=-a_{0}a_{1}, \\
&2a_{7}-2a_{2}+\frac{1}{2}a_{3}+\frac{1}{2}a_{9}=a_{0}-a_{0}a_{2}+\frac{1}{2}a_{0}^{2}, \\ &2a_{8}+2a_{1}-\frac{3}{2}a_{3}+a_{6}=a_{0}-a_{0}a_{3}+\frac{1}{2}a_{0}^{2}.\label{c-8}
\end{align}
A simple solution is obtained for the choice of parameters $a_{0}=a_{1}=a_{8}=0$. In that case from \eqref{c-1}-\eqref{c-8} we find $a_{2}=0,\;a_{3}=1,\; a_{4}=0,\; a_{5}=\frac{1}{2}, \; a_{6}=\frac{3}{2}, \; a_{7}=0$ and $a_{9}=-1$, i.e.~we have that
\begin{equation}
\tilde{f}=1-\frac{1}{2}z-\frac{1}{4}v^{2}-\frac{1}{4}z^{2}+vz,
\end{equation}
and
\begin{equation}
\tilde{g}=z+\frac{1}{2}v^{2}+\frac{3}{2}z^{2}-vz,
\end{equation}
i.e.~$\hat p_\mu = p_\mu \tilde f + (\alpha/\beta) x_\mu \tilde g$.

\section{Conclusions}
TSR deforms relativistic quantum mechanics by two fundamental parameters, and is related to a model proposed long time ago by C.N.~Yang.

In this paper, we have proposed a class of non linear algebras satisfying the Jacobi identities that generalize TSR, and discussed their realizations in terms of the Heisenberg algebra.
Note that the methods proposed in \cite{Meljanac-3} for Lie-algebra type models cannot be applied.

We have also considered the classical limit obtained by replacing commutators by Poisson brackets and discussed in detail its realizations on canonical phase space.
A different approach to the representation of these models would be to realize the algebra on an extended phase space \cite{Martinic-1}.

It is not clear whether is it possible to generalize the coalgebra structure, addition of momenta, star product
and twist for these non linear algebra models.

It would be interesting to discuss the physical implications of these generalized models, including their nonrelativistic limit, for example on
the deformation of the uncertainty relations or on some simple systems, as in \cite{Mignemi-2}. A more ambitious task would be the construction of a quantum
field theory in this framework, but this would require a departure from the standard formalism.

\section*{Acknowledgements}
S. Mignemi acknowledges support of Gruppo Nazionale di Fisica Matematica

\end{document}